**EFECTO DE UNA MEZCLA DE AMOXICILINA Y NORFLOXACINA EN LA ALIMENTACIÓN DE LECHONES EN RECRÍA SOBRE EL COMPORTAMIENTO PRODUCTIVO Y LA INCIDENCIA DE CUADROS CLÍNICOS**

**EFFECT OF A MIXTURE OF AMOXICILLIN AND NORFLOXACIN IN PIGLETS FEEDING ON PRODUCTIVE PERFORMANCE AND CLINICAL SIGNS**

Diego Martínez[1], Enrique Alvarado[2] y Carlos Vílchez[3]

**ABSTRACT**

An experiment was conducted to evaluate the effect of feeding post-weaned piglets different levels of a commercial amoxicillin plus norfloxacin formula (Respirend®) on the performance, clinical evidence of sickness, and the cost:benefit ratio. 454 PIC x Camborough female and castrated male, weaned at 21 d of age, with an average initial weight of 6.19 kg, were used in this experiment. Treatment 1 consisted in feeding a basal diet supplemented with 1000 g of a commercial tilmicosin product (Pulmotil®) per metric ton of feed (control group, from 21 to 35 d of age) and Treatments 2), 3) and 4) consisted in feeding the basal diet supplement with 500, 300 and 100 g of Respirend® per metric ton of feed, respectively, from 21 to 42 d of age. The results did not show significant differences (P>0.05) on mortality nor on parenteral medicine administration cost in none of the periods. Nevertheless, 500 g of Respirend® per metric ton of feed improved significantly (P<0.05) the general health status in the period between 21 and 35 d of age. For the whole evaluation period (21 to 42 d of age), this level improved significantly (P<0.05) the final body weight, the average daily weight gain, the average daily feed intake, the feed:gain ratio and optimized the cost:benefit ratio.

**RESUMEN**

Se ha llevado a cabo un experimento para evaluar el efecto de la inclusión de diferentes niveles de una fórmula comercial de amoxicilina y norfloxacina (Respirend®) en la dieta de lechones destetados sobre el comportamiento productivo, la incidencia de cuadros clinicos y la relación costo-beneficio. Se utilizaron 454 lechones machos castrados y hembras

---

[1] Ingeniero Zootecnista, Práctica Privada.
[2] Profesor Principal, Departamento de Producción Animal - UNALM.
[3] Profesor Principal, Departamento de Nutrición - UNALM.

EFECTO DE UNA MEZCLA DE AMOXICILINA Y NORFLOXACINA EN LA ALIMENTACIÓN DE LECHONES EN RECRIA SOBRE EL COMPORTAMIENTO PRODUCTIVO Y LA INCIDENCIA DE CUADROS CLINICOS

de la linea PIC- Camborough, destetados a los 21 días de edad y con 6.19 kg de peso promedio. El tratamiento 1 (control) recibió un producto comercial en base a tilmicosina (Pulmotil®) a razón de 1000 g/t de alimento (de 21 a 35 días de edad) y los tratamientos 2,3 y 4 recibieron Respirend® a razón de 500, 300 y 100 g/t de alimento, respectivamente, de 21 a 42 días de edad. Para el análisis de los datos se utilizó el Diseño de Bloques Completo al Azar. Los resultados del estudio no mostraron diferencias significativas (P>0.05) en la mortalidad ni en los costos por concepto de aplicación parenteral de medicinas en ninguno de los periodos evaluados. Sin embargo, el nivel de 500 g de Respirend® por tonelada de alimento mejoró significativamente (P<0.05) el estado general de salud en el período de 21 a 35 días de edad. En el período total de evaluación (de 21 a 42 días de edad) este nivel mejoró significativamente (P<0.05) el peso final, la ganancia de peso, el consumo de alimento, el índice de conversión del alimento y la relación costo:beneficio.

**Palabras claves:** Amoxicilina, Norfloxacina, Tilmicosina, Antibióticos, Lechones

## INTRODUCCION

El período alrededor del destete es un momento crítico de la vida del cerdo joven. El cambio de la leche por alimento sólido, la falta de un sistema inmune completamente desarrollado, el cambio de ambiente y de jerarquía social ponen en peligro la salud del animal. Con la finalidad de reducir el efecto del estrés sobre la salud del animal durante esta transición se utilizan antibióticos en el alimento. Este hecho hace posible la evaluación constante de nuevos productos que permitan obtener mejores rendimientos, favorezcan la salud del animal y permitan reducir los costos de producción.

El objetivo del presente estudio fue evaluar el efecto de la inclusión de diferentes niveles de una asociación antimicrobiana compuesta por amoxicilina y norfloxacina (Respirend®) en la dieta de lechones destetados sobre el comportamiento productivo, la incidencia de cuadros clínicos y la relación costo-beneficio.

## REVISIÓN DE LITERATURA

El uso de antibióticos como aditivos del alimento en la industria ganadera ha resultado en enormes ahorros al mejorar la tasa de ganancia de peso y la conversión alimenticia, reduciendo la morbilidad y la mortalidad resultantes de las infecciones clínicas o sub clínicas (Beltranena y Aherne, 1991) e incrementando la calidad de los productos finales.

Los antibióticos ejercen cierto efecto sobre el metabolismo animal, incluyendo: eliminación de la depresión del crecimiento por productos metabólicos, mejor aprovechamiento de aminoácidos, aumento celular y proteico, y la activación de las funciones suprarrenales y de la tiroides. Asimismo, reducen la hidrólisis de urea y ácidos biliares (Necochea, 1982).

El «factor de crecimiento similar a la insulina I» (IGF-I) que es una proteína que media varios efectos de la hormona del crecimiento y promueve el crecimiento de tejido magro, y la «glicoproteína ácida 0.1» (AGP) que es una proteína de la fase aguda que trabaja retroalimentando negativamente al sistema inmune y reduciendo la respuesta inflamatoria.

Los antimicrobianos incrementan los niveles séricos de IGF-I y reducen los de AGP (Weber *et al.*, 1999).

Los antibióticos inhiben microorganismos que, de otra manera, podrían producir enfermedades subclínicas retardando el crecimiento. Asimismo, los microorganismos gastrointestinales compiten con su hospedero por los nutrientes de la dieta y pueden tener un efecto nocivo sobre el desempeño de los pollos de engorde, influyendo en los requerimientos nutritivos, afectando la morfología del tracto intestinal y modificando la actividad metabólica (Szylit y Charlet, 1981).

Ciertos antibióticos reducen el espesor de la pared intestinal, lo que resulta en una potencial mayor absorción de nutrientes. En el caso de compuestos inorgánicos, la relación inversamente proporcional entre el grado de absorción y el grosor intestinal puede ser una causa y efecto directos; mientras que en el caso de compuestos orgánicos puede deberse a la combinación del engrosamiento y la oxidación microbiana (McAllister *et al.*, 1979).

La edad del cerdo es determinante en la respuesta a los antibióticos. Los incrementos en la ganancia diaria de peso promedio, resultante del uso de antibióticos en las fases de inicio y crecimiento-engorde son de 15 y 3.6%, respectivamente; mientras que el uso de antibióticos afecta la conversión alimenticia de las mismas fases mejorándolas en 6.5 y 2.4%, respectivamente. Se ha demostrado que las mejoras en los índices productivos son menores cuando el estado de salud del lote y las instalaciones presentan condiciones de buenas a excelentes (Reese *et al.*, 2000).

La amoxicilina es un antibiótico semisintético con un amplio espectro de actividad bactericida, que inhibe la síntesis de pared celular bacteriana. Su absorción tras ser administrada oralmente no es afectada por el alimento. La mayor parte de la amoxicilina es excretada en la orina la norfloxacina es una fluoroquinolona de origen sintético con un amplio espectro de acción bactericida a través de la inhibición de la encima ADN-girasa. La administración simultánea con el alimento tiene un efecto despreciable sobre la absorción del antibiótico (Wise, 1984). Es importante tener en consideración que la utilización indiscriminada de algunos antibióticos en explotaciones pecuarias es un factor predisponente para la aparición de cepas resistentes (Prescott, 2000).

**MATERIALES Y MÉTODOS**

El experimento se llevó a cabo en las instalaciones de la empresa Agroindustria Campoy S.A.C., Lima-Perú, evaluándose un periodo de tres semanas a partir del destete. Se trabajó con 454 lechones destetados machos castrados y hembras de la línea PIC-Camborough, con 21 días de edad y 6.19 kg de peso promedio, que fueron agrupados en 6 lotes y alojados a razón de 0.25 m2. Se utilizó una mezcla de amoxicilina (15%) y norfloxacina (10%) (Respirend®) y una tilmicosina (10%) (Pulmotil®). La tilmicosina es un antibiótico macrólido semisintético. Se evaluaron cuatro tratamientos: TI, medicado con Pulmotil® a razón de 1000 g/t de alimento (control, de 21 a 35 días de edad) y T2, T3 y T4 medicados con Respirend®, de 21 a 42 días de edad, a razón de 500, 300 y 100 g/t de alimento, respectivamente. Se suministraron ad libitum dos tipos de alimentos, elaborados a base a maíz y harina de pescado: Fase I (de 21 a 35 días de edad) con 22% de PT y 3497 Kcal/kg EM y Fase II (de 36 a 42 días de edad) con 21% de PT y 3435 Kcal/Kg EM. La composición porcentual de las dietas basales, tanto de la Fase I como de la Fase 11, se presenta en el Cuadro 1. Los parámetros evaluados fueron: El peso vivo (kg), la ganancia de peso 7 (g/animal/día), el consumo de alimento (g/animal/día), la conversión alimenticia, los días-animal del total de enfermedades (días/animal), la mortalidad (%), los costos terapéuticos adicionales (US $/Kg de peso ganado) y la relación costo-beneficio. Se utilizó un Diseño de Bloques Completo al Azar con cuatro tratamientos y seis repeticiones por tratamiento. El análisis de varianza de los datos se llevó a cabo usando el programa Statistical Análisis System (SAS) y la diferencia de medias se determinó mediante la prueba de Duncan.

## RESULTADOS Y DISCUSIÓN

### Periodo experimental de 21 a 35 días

En el Cuadro 2 se detallan los resultados del presente estudio. El peso vivo (PV) no mostró diferencias estadísticas significativas (DES) (P>0.05) análisis de varianza (ANVA) a los 21 días, ni a los 35 días de edad; sin embargo, al día 35, los animales medicados con Respirend® mostraron PV numéricamente mayores con respecto al control en 7.71, 4.28 y 2.35% con 500, 300 y 100 g/t, respectivamente. Esto podría indicar que el tiempo de evaluación transcurrido no ha sido suficiente para que los animales manifiesten el efecto de los tratamientos.

La ganancia diaria de peso (GDP) no mostró DES (P>0.05), debido posiblemente a que los antimicrobianos no eliminaron en magnitudes diferentes los organismos productores de toxinas como para incrementar la absorción de nutrientes (Hamilton y Proudfoot, 1991). Se aprecia una tendencia positiva para los tratamientos medicados con Respirend® con respecto al tratamiento medicado con Pulmotil® (GDP 25.84, 15.26 y 9.64% mayores con 500, 300 y 100 g/t, respectivamente). Se ha demostrado que dosis profilácticas de antibióticos más elevadas tienen como resultado mayores GDP (Herrmann, 1978; citado por Girano, 1988), aunque la diferencia entre las ganancias no sea significativa.

El consumo de alimento no mostró DES (P>0.05). Esto concuerda con lo reportado por Lindemann *et al*. (1985), quienes no observaron DES en el consumo de alimento de los cerdos tras suministrar cuatro niveles de salinomicina por un período de cuatro semanas.

La conversión alimenticia no mostró DES (P>0.05) al ANVA; sin embargo, con 500 g/t de Respirend® se obtuvo un índice significativamente (P<0.05) mejor que en el grupo control (20.07, 14.79 y 13.98%, con 500, 300 y 100 g/t, respectivamente). En esta etapa, la mejor conversión alimenticia observada con 500 g/t de Respirend® no se relacionó en términos significativos con la GDP ni con el consumo de alimento; sin embargo, es importante recordar que la conversión alimenticia es resultado de la interacción de estos parámetros.

Los días-animal del total de enfermedades (DATE) no mostraron DES (P>0.05) al ANVA. Los animales que recibieron 500 g/t de Respirend® en el periodo de 21 a 42 días. Resulta difícil establecer la diferencia entre el efecto directo sobre el comportamiento productivo y la supresión de la infección definida cuando se suministran antibióticos en el alimento. El efecto puede atribuirse a un mayor consumo de alimento y también a la reducción de los procesos entéricos (Lucas y Lodge, 1967). Los cuadros clínicos observados fueron principalmente consecuencia de procesos entéricos, epidermitis exudativa, dermatitis auricular necrótica, entre otros. La mortalidad fue muy baja (+1.75%) y no estuvo asociada a los tratamientos dietarios. Los costos terapéuticos adicionales (CTA) no mostraron DES (P>0.05). Sin embargo, se observa una tendencia positiva congruente con la observada en los DATE.

Con 500, 300 y 100 g/t de Respirend®, se observaron relaciones costo-beneficio (RCB) 20.18, 16.47 y 16.49%, respectivamente, más eficientes que con el tratamiento control.

**Período experimental de 35 a 42 días**

La medicación con 500 g/t de Respirend® promovió un mayor PV (P<0.05) que 1000 g/t de Pulmotil® al día 42 edad. Las DES (P<0.05) en los pesos a los 42 días, y su tendencia constante en el tiempo, demuestran que el tiempo de evaluación transcurrido hasta los 35 días de edad fue insuficiente para que el efecto de los tratamientos pudiera ser expresado a niveles significativos. Es necesario tener en cuenta que el retiro de la medicación del grupo control pudo favorecer a los demás tratamientos.

Los animales del grupo control mostraron la menor GDP (P<0.05). Se observaron GDP 25.50, 24.83 y 14.72% mayores que el control con 500, 300 y 100 g/t de Respirend®, respectivamente. Los animales que recibieron 500 g/t de Respirend® consumieron significativamente (P<0.05) más alimento que el grupo control. El tratamiento medicado con 100 g/t de Respirend® muestra un consumo de alimento 2.94% menor que el control. Esto posiblemente esté relacionado con el estado de salud en este periodo.

La conversión alimenticia no mostró DES (P>0.05) al ANVA. Sin embargo, según la prueba de Duncan, los índices de conversión del alimento fueron 20.21, 19.90 y 21.70% más eficientes (P<0.05), con 500, 300 y 100 g/t de Respirend®, respectivamente.

Los DATE no mostraron DES (P>0.05). Los animales que recibieron 500, 300 y 100 g/t de Respirend® mostraron estar enfermos 32.36, 5.11 y 46.96% menos días-animal, respectivamente. En este periodo, con 100 g/t de Respirend® se observó un menor consumo de alimento (P>0.05) en relación al grupo control. Estos mismos animales muestran una mejora en su estado de salud en relación al periodo anterior con respecto al grupo control. Al presentar un mejor estado de salud, estos animales podrían haber requerido una menor cantidad de nutrientes para soportar su sistema inmune. Williams *et al*. (2001) indican que el efecto neto de un reto inmunológico aparenta ser redistribución de nutrientes hacia procesos diferentes al crecimiento como soportar el sistema inmune. La mortalidad fue nula en todos los tratamientos. No se observaron DES (P<0.05) en los CTA. Se observó, en términos generales, un menor requerimiento de medicinas en este periodo en relación al período anterior.

Con 500, 300 y 100 g/t de Respirend®, se observaron RCB 33.33, 44.00 y 54.67%, respectivamente, más eficientes que con el tratamiento control.

**Período experimental de 21 a 42 días**

La GDP fue significativamente (P<0.05) menor con 1000 g/t de Pulmotil® en comparación a 500 ó 300 g/t de Respirend®. La inclusión de 50,300 y 100 g/t Respirend® mejoró la GDP en 23.86, 18.18 y 10.93%, respectivamente. En esta etapa, Respirend® a 500 g/t mejoró (P<0.05) la GDP respecto al control y mostró un consumo de alimento 5.5% mayor, además de un mejor (P<0.05) estado de salud en el período inicial de evaluación.

EFECTO DE UNA MEZCLA DE AMOXICILINA Y NORFLOXACINA EN LA ALIMENTACIÓN
DE LECHONES EN RECRIA SOBRE EL COMPORTAMIENTO PRODUCTIVO
Y LA INCIDENCIA DE CUADROS CLINICOS

El consumo de alimento no mostró DES (P>0.05). Sólo los animales que recibieron 100 g/t de Respirend® consumieron menos alimento que el grupo control (1.38% menos).

Los tres niveles de Respirend® convirtieron el alimento más eficientemente (P<0.05) que el grupo control, requiriéndose 14.48, 11.77 y 11.43% menos alimentos por kg de peso ganado con 500, 300 y 100 g/t, respectivamente. Las conversiones del alimento, así como las tasas de crecimiento observadas con Respirend® fueron más favorables que con el tratamiento con Pulmotil®; sin embargo, los incrementos en el consumo de alimento, a pesar de ser significativos, fueron en términos relativos menores a los observados en ganancia de peso. The Council for Agricultural Science and Technology (CAST) (1981), citado por Lindemann *et al*. (1985), resume 203 experimentos con cerdos en recría alimentados con dietas suplementadas con antibióticos. Con el 90% de los antibióticos evaluados, se obtuvo como tendencia que, a diferencia de lo observado en el presente estudio, los incrementos en las tasas de GDP ocurren como resultado del incremento en el consumo de alimento y de la mejora de la conversión alimenticia.

Los DATE no mostraron DES (P>0.05); sin embargo, la tendencia observada en el período de 21 a 35 días de edad, definitivamente se sigue manteniendo. La mortalidad fue muy baja (+1.75%) y no estuvo asociada con los tratamientos dietarios. No se observaron DES (P>0.05) en los CTA.

Los animales medicados con 500,30 y 100 g/t de Respirend®, mostraron RCB 54.35,47.83 y 52.17%, respectivamente, más eficientes que los medicados por el tratamiento control, como resultado directo de la utilización más eficiente del alimento.

## CONCLUSIONES

Los resultados obtenidos bajo las condiciones del presente estudio permiten concluir que la inclusión de la fórmula comercial de amoxicilina y norfloxacina (Respirend®) en la dieta de lechones destetados, a razón de 500 y 300 9 por tonelada de alimento, mejora (P<0.05) la ganancia de peso y la conversión alimenticia. Además, se encontró que la inclusión de 500 g Respirend® por tonelada de alimento optimizó la relación costo:beneticio del alimento.

## AGRADECIMIENTO

## BIBLIOGRAFIA

EFECTO DE UNA MEZCLA DE AMOXICILINA Y NORFLOXACINA EN LA ALIMENTACIÓN
DE LECHONES EN RECRIA SOBRE EL COMPORTAMIENTO PRODUCTIVO
Y LA INCIDENCIA DE CUADROS CLINICOS

**Cuadro 1. Composición de las dietas basales Fase I y Fase II.**

| Ingredientes, % | Dietas basales | |
| --- | --- | --- |
| | Fase I | Fase II |
| Maíz | 27.57 | 60.53 |
| Harina de pescado prime (67% Pt) | 19.34 | 18.62 |
| Suero de leche desecado | 15.47 | 7.77 |
| Suero de leche proteinizado en polvo | 9.92 | 4.98 |
| Polvillo de arroz | 5.95 | 6.67 |
| Azúcar rubia | 9.92 | - |
| Avena | 9.92 | - |
| Aceite de palma | - | 0.30 |
| Carbonato de calcio | 0.69 | 0.50 |
| Fosfato monodicálcico | 0.50 | 0.30 |
| Premezcla de inicio[1] | 0.20 | 0.15 |
| Secuestrante de micotoxinas | 0.05 | - |
| Oxido de zinc | 0.30 | 0.20 |
| Acid Pack® | 0.20 | - |
| **Total** | **100.00** | **100.00** |
| *Contenido nutricional (calculado)* | | |
| Proteína cruda, % | 22.0 | 21.0 |
| Energía metabolizable, kcal/kg | 3497.2 | 3435.0 |
| Lisina, % | 1.62 | 1.40 |
| Metionina, % | 0.57 | 0.56 |
| Metionina + cisteína, % | 0.86 | 0.83 |
| Treonina, % | 1.03 | 0.95 |
| Triptófano, % | 0.28 | 0.25 |
| Fibra cruda, % | 2.0 | 2.5 |
| Calcio, % | 1.00 | 0.88 |
| Fósforo disponible, % | 0.53 | 0.49 |
| Grasa total, % | 4.90 | 5.31 |
| Cobre, ppm | 9.26 | 4.94 |
| Zinc, ppm | 2017 | 1534 |
| Selenio, ppm | 0.42 | 0.41 |
| Ac. linoléico, % | 0.86 | 1.61 |
| Ac. fólico, mg/kg | 0.46 | 0.42 |

Composición por kg de premezcla: Vitamina A: 9´000,000 U.I.; Vitamina D3: 1´000,000 U.I.; Vitamina E: 45,000 U.I.; Niacina: 20 g; Acido pantoténico: 15 g; Vitamina K: 3 g; Riboflavina: 2.5 g; Tiamina: 1 g; Acido fólico: 0.5 g; Biotina 0.25 g; Vitamina B12: 0.018 g; Hierro: 80 g; Zinc: 50 g; Manganeso: 40 g; Cobre: 5 g; Yodo: 1.5 g; Selenio: 0.1 g.

**Cuadro 2: Comportamiento productivo y condiciones clínicas observadas en lechones destetados alimentados con dietas conteniendo Pulmotil® y diferentes niveles de Respirend®**

| Parámetro | Tratamiento* | | | | | | | |
|---|---|---|---|---|---|---|---|---|
| | 1 | | 2 | | 3 | | 4 | |
| **Etapa de 21 a 35 días de edad** | | | | | | | | |
| Peso Vivo Inicial (día 21), kg | 6.22 | | 6.23 | | 6.21 | | 6.18 | |
| Peso Vivo Final (día 35), kg | 8.83 | | 9.51 | | 9.20 | | 9.03 | |
| Ganancia de Peso, g/día | 173 | | 218 | | 200 | | 190 | |
| Consumo de Alimento, g/día | 359 | | 369 | | 370 | | 359 | |
| Conversión Alimenticia | 2.217 | b | 1.772 | a | 1.889 | ab | 1.907 | ab |
| Días-Animal del Total de Enfermedades[1], d/a | 1.45 | b | 0.84 | a | 1.03 | ab | 1.02 | ab |
| Mortalidad, % | 0.00 | | 0.00 | | 0.88 | | 1.75 | |
| Costos Terapéuticos Adicionales[2], US.$/kg | 0.049 | | 0.029 | | 0.034 | | 0.040 | |
| Relación Beneficio-Costo[3], US $ | 0.25 | | 0.53 | | 0.42 | | 0.42 | |
| **Etapa de 35 a 42 días de edad** | | | | | | | | |
| Peso Vivo Inicial (día 35), kg | 8.83 | | 9.51 | | 9.20 | | 9.03 | |
| Peso Vivo Final (día 42), kg | 11.25 | c | 12.45 | a | 12.14 | ab | 11.75 | bc |
| Ganancia de Peso, g/día | 336 | b | 422 | a | 419 | a | 385 | a |
| Consumo de Alimento, g/día | 706 | bc | 765 | a | 753 | ab | 686 | c |
| Conversión Alimenticia | 2.286 | b | 1.824 | a | 1.831 | a | 1.790 | a |
| Días-Animal del Total de Enfermedades[1], d/a | 0.41 | | 0.28 | | 0.39 | | 0.22 | |
| Mortalidad, % | 0 | | 0 | | 0 | | 0 | |
| Costos Terapéuticos Adicionales[2], US.$/kg | 0.001 | | 0.002 | | 0.003 | | 0.002 | |
| Relación Beneficio-Costo[3], US $ | 0.75 | | 1.00 | | 1.08 | | 1.16 | |
| **Etapa de 21 a 42 días de edad** | | | | | | | | |
| Peso Vivo Inicial (día 21), kg | 6.22 | | 6.23 | | 6.21 | | 6.18 | |
| Peso Vivo Final (día 42), kg | 11.25 | c | 12.45 | a | 12.14 | ab | 11.75 | bc |
| Ganancia de Peso, g/día | 228 | c | 283 | a | 270 | ab | 253 | bc |
| Consumo de Alimento, g/día | 469 | ab | 495 | a | 491 | a | 463 | b |
| Conversión Alimenticia | 2.071 | b | 1.771 | a | 1.827 | a | 1.834 | a |
| Días-Animal del Total de Enfermedades[1], d/a | 1.86 | | 1.12 | | 1.42 | | 1.24 | |
| Mortalidad, % | 0.00 | | 0.00 | | 0.88 | | 1.75 | |
| Costos Terapéuticos Adicionales[2], US.$/kg | 0.023 | | 0.015 | | 0.018 | | 0.021 | |
| Relación Beneficio-Costo[3], US $ | 0.46 | | 0.71 | | 0.68 | | 0.70 | |

\* Tratamientos 1: Control Pulmotil® a 1000 g/t (de 21 - 35 días de edad; 35 - 42 días de edad, sin antibiótico); 2: Respirend® a 500 g/t; 3: Respirend® a 300 g/t; 4 : Respirend® a 100 g/t.

[1] Se define como la suma de las duraciones de cada uno de los cuadros clínicos observados en cada uno de los animales afectados por cualquiera de ellos.
[2] Se define como la suma de los costos incurridos por concepto de aplicación parenteral de medicinas.
[3] Indica el monto neto ganado, en peso vivo, por cada dólar invertido en alimento.

a, b, c: Promedios dentro de una misma línea con diferentes letras son diferentes significativamente (P<0.05)